\documentclass[traditabstract]{aa} % for the abstract without structuration
\usepackage{graphicx}
\usepackage{txfonts}
\usepackage{natbib}
\usepackage{color}
\usepackage{lscape}
\usepackage{latexsym}
\usepackage{changepage}		%for adjustwidth
\bibpunct{(}{)}{;}{a}{}{,} % to follow the A&A style
\usepackage[usenames,dvipsnames,svgnames,table]{xcolor}
\usepackage[breaklinks, colorlinks, citecolor=CornflowerBlue]{hyperref}

\begin{document}
\definecolor{Red}{rgb}{1,0,0}
\authorrunning{Remco F.J. van der Burg, et al.}
   \title{The abundance and spatial distribution of ultra-diffuse galaxies in nearby galaxy clusters}
   \author{Remco~F.J.~van der Burg\inst{1}, Adam~Muzzin\inst{2}, and Henk~Hoekstra\inst{3}}
	   \institute{Laboratoire AIM, IRFU/Service d'Astrophysique - CEA/DSM - CNRS - Universit\'e Paris Diderot, B\^at. 709, CEA-Saclay, 91191 Gif-sur-Yvette Cedex, France
		\and Kavli Institute for Cosmology, University of Cambridge, Madingley Road, Cambridge, CB3 0HA, United Kingdom
	      \and Leiden Observatory, Leiden University, P.O. Box 9513, 2300 RA Leiden, The Netherlands\\
                    \email{remco.van-der-burg@cea.fr}        
        }             
             
   \date{Submitted 29 January 2016 ; Accepted 24 March 2016}
% \abstract{}{}{}{}{} 
% 5 {} token are mandatory

  \abstract{Recent observations have highlighted a significant population of faint but large ($\mathrm{r_{eff}>1.5\,kpc}$) galaxies in the Coma cluster. The origin of these Ultra Diffuse Galaxies (UDGs) remains puzzling, as the interpretation of these observational results has been hindered by the (partly) subjective selection of UDGs, and the limited study of only the Coma (and some examples in the Virgo-) cluster.
   In this paper we extend the study of UDGs using eight clusters in the redshift range $0.044<z<0.063$ with deep $g$- and $r$-band imaging data taken with MegaCam at the CFHT. We describe an automatic selection pipeline for quantitative identification, and tested for completeness using image simulations of these galaxies. We find that the abundance of the UDGs we can detect increases with cluster mass, reaching $\sim 200$ in typical haloes of $M_{200}\simeq 10^{15}\,\mathrm{M_{\odot}}$. For the ensemble cluster we measure the size distribution of UDGs, their colour-magnitude distribution, and their completeness-corrected radial density distribution within the clusters. The morphologically-selected cluster UDGs have colours consistent with the cluster red sequence, and have a steep size distribution that, at a given surface brightness, declines as $\mathrm{n [dex^{-1}]\propto r_{eff}^{-3.4\pm0.2}}$. 
   Their radial distribution is significantly steeper than NFW in the outskirts, and is significantly shallower in the inner parts. We find them to follow the same radial distribution as the more massive quiescent galaxies in the clusters, except within the core region of $r \lesssim 0.15 \times R_{200}$ (or $\lesssim 300$ kpc). Within this region the number density of UDGs drops and is consistent with zero. These diffuse galaxies can only resist tidal forces down to this cluster-centric distance if they are highly centrally dark-matter dominated. The observation that the radial distribution of more compact dwarf galaxies ($\mathrm{r_{eff}<1.0\,kpc}$) with similar luminosities follows the same distribution as the UDGs, but exist down to a smaller distance of 100 kpc from the cluster centres, may indicate that they have similarly massive sub-haloes as the UDGs. Although a number of scenarios can give rise to the UDG population, our results point to differences in the formation history as the most plausible explanation.}
   \maketitle
%
%________________________________________________________________

\hyphenation{in-tra-clus-ter}
\hyphenation{rank-or-der}

\section{Introduction}
There is a great interest in studying the dwarf galaxy population in the Universe. Not only do these low-luminosity galaxies dominate the total number density distribution of galaxies \citep[e.g.][]{blanton03,loveday12}, but they also provide sensitive constraints to complex feedback processes that regulate the formation of stars on these scales \citep{weinmann12}. Dwarf galaxies exist comprising a range of physical properties that is even more diverse than seen in recent hydrodynamical simulations \citep{oman15}. It is an observational challenge to study the dwarf galaxy population far beyond the local group, especially when their stars are spread into diffuse systems. 

Recognizing the potential of galaxies with low surface brightnesses to constrain complex physical phenomena, several past studies \citep[e.g.][]{sandage84,impey88,turner93,dalcanton97} have studied galaxies at the lowest surface brightnesses observable at that time. These early studies already indicated a large diversity in the properties of the dwarf galaxy population, along with the occurrence of faint extended galaxies, both in the field \citep{dalcanton97}, and in cluster-like environments \citep{sandage84,impey88,turner93}. With the technological advancement of sensitive CCD cameras with a larger field-of-view, these studies can now be performed to increasingly fainter limits. 

One of these new instruments is the Dragonfly Telescope Array \citep{dragonfly}, which allows for a detailed study of low surface brightness features due to a clean optical system without significant internal reflections. While studying extended low-surface-brightness patterns around the Coma cluster using this camera array, \citet{vandokkum15} encountered a population of large galaxies ($\mathrm{r_{eff}\sim3-10''}$) with relatively low central surface brightnesses ($\mu(g,0)=24-26\,\mathrm{mag\,arcsec^{-2}}$); these may thus be a fainter extension of the population studied in the aforementioned works. Given that the projected density of these galaxies coincides with the Coma cluster, and has a significant overdensity compared to the surrounding fields, \citet{vandokkum15} concluded that these are likely part of the Coma cluster. They introduced the term Ultra Diffuse Galaxies (UDGs) for these galaxies with effective radii $\mathrm{r_{eff}>1.5\, kpc}$, and surface brightnesses in the range mentioned above. A follow-up study spectroscopically confirmed one of the UDGs to be at the redshift of the Coma cluster \citep{vandokkum15b}. 

Since then, several other studies have reported the discoveries of more galaxies with similar properties. While studying deep Suprime-Cam data taken with the Subaru telescope, and exploring a slightly larger region of parameter space (size versus surface brightness), \citet{koda15} estimated the existence of $\sim$1000 diffuse galaxies in the Coma cluster. \citet{mihos15} identified and studied three galaxies with even lower surface brightnesses in the Virgo cluster.  

With their existence firmly established, it is an open question how such diffuse galaxies can exist in these highly over-dense environments. Galaxies with stellar masses as low as that of these UDGs are generally blue and star forming in the field \citep[e.g.][]{davies16}, although the effective volume from which such faint galaxies can be studied in deep enough surveys is limited. Also in the Virgo cluster a population of faint blue low surface brightness galaxies was already identified some decades ago \citep{impey88}. The red colours of the newly discovered population in clusters, on the contrary, suggest that they have been quenched some time ago, likely around redshift $z\simeq 2$. If this quenching process happened while these galaxies fell into the cluster, the question is how these galaxies have been able to survive its harsh environment for several Gyr. \citet{vandokkum15} speculated that they are surrounded by relatively massive dark matter subhaloes, which would make them relatively resistant to tidal forces from the cluster potential. Only in the central 300 kpc would the tidal forces be strong enough to lead to the dissociation of the UDG, explaining that none of them are found this close to the cluster centres. 

\citet{yozinbekki15} performed a numerical simulation to test a scenario in which these galaxies are accreted at early times onto the cluster and are quenched by ram pressure stripping during first infall. Within the limits of the simulation, they can already reproduce some of the observed properties of the UDGs, such as their colour, their size, their disk-like morphology, and their absence in the central parts of the cluster. To be able to fully understand the origin of these galaxies, simulations have to be expanded by including additional physical processes like harassment by other cluster members \citep[e.g.][]{Moore1996}, which is currently missing in this simulation but may also have a significant effect on the survival of UDGs in the clusters. 

A current limitation for the models and simulations is that the underlying distribution of galaxies, from which the observed samples of UDGs are selected, is poorly understood. The studied samples are (at least partly) selected by-eye \citep{vandokkum15,koda15,mihos15}, which renders the samples difficult to reproduce, and a quantitative analysis difficult. Furthermore, it is still unclear if the galaxies found in Coma are a general phenomenon for clusters in the local Universe, or whether they are particular to this one studied cluster. 

To help address these points, this paper studies UDGs in eight clusters with deep $g$- and $r$-band data taken with MegaCam@CFHT as part of the Multi-Epoch Nearby Cluster Survey \citep[MENeaCS, e.g.][]{sand11}. We will define quantitative selection criteria to select these UDGs using standard software (\texttt{SExtractor} and \texttt{GALFIT}) in the clusters. Moreover, we use data from the CFHT Legacy Survey to perform a statistical background correction. We measure several of their properties, such as their colours, stellar masses, size distributions, and their radial distribution in the clusters. To better understand the origin and evolution of the UDGs, we will also compare their radial distribution to that of more typical dwarf galaxies with similar luminosities but more compact morphologies.

The paper is organised as follows. Section~\ref{sec:dataoverview} provides an overview of the data products we use, Sect~\ref{sec:analysis} describes how we select our sample of UDGs. The properties of UDGs are described in the following three sections: Their colours and stellar masses in Sect.~\ref{sec:colours}, their abundances and size distributions in Sect.~\ref{sec:abundancehalomass}, and their radial distribution in the clusters in Sect.~\ref{sec:radialdist}. We discuss our findings in Sect.~\ref{sec:discussion}, and summarize in Sect.~\ref{sec:conclusion}.

All magnitudes we quote are in the AB magnitudes system, and we adopt angular-diameter and luminosity- distances corresponding to $\Lambda$CDM cosmology with $\Omega_{\mathrm{m}}=0.3$, $\Omega_{\Lambda}=0.7$ and $\mathrm{H_0=70\, km\, s^{-1}\,  Mpc^{-1}}$. For stellar masses we assume the Initial Mass Function (IMF) from \citet{chabrier03}.

\section{Data overview \& processing}\label{sec:dataoverview}
The sample we study is drawn from the Multi-Epoch Nearby Cluster Survey (MENeaCS), which is an optical follow-up survey of X-ray luminous clusters in the local Universe ($0.05<z<0.15$), using MegaCam at the CFHT. This paper focuses on eight nearby MENeaCS clusters with redshifts $z<0.07$, ensuring that UDGs stand out most significantly against the background. Out of the 10 clusters that satisfy this redshift criterion \citep[][hereafter vdB15]{vdB15}, the eight were selected to be at high Galactic latitude $|$\textit{b}$_{\mathrm{Gal}}|>25\degr$ to reduce obscuration from stars in the Galaxy, see Table~\ref{tab:overview}. Note that A553 and A2319 are at $|$\textit{b}$_{\mathrm{Gal}}|<15\degr$, and are not considered for this study. The photometry is described in detail in vdB15, and some relevant information for the present study is summarized below.

\begin{table*}%[ht]
\caption{The cluster sample studied here.}
\label{tab:overview}
\begin{center}
%\begin{adjustwidth}{-0.4cm}{}
\begin{tabular}{l c r r r cc}
\hline
\hline
Cluster & $z_{\mathrm{spec}}$ & RA & Dec & \textit{b}$_{\mathrm{Gal}}$ &$M_{\mathrm{200}}^{\mathrm{a}}$ & $R_{\mathrm{200}}^{\mathrm{a}}$ \\
&&J2000&J2000&[$\degr$]&[$10^{14}\,\mathrm{M_{\odot}}$]&[Mpc]\\
\hline
       A85&0.055&00:41:50.33&-09:18:11.20&-72.0&$ 10.0\pm  1.7$&$  2.0\pm  0.1$\\
      A119&0.044&00:56:16.04&-01:15:18.22&-64.1&$\,\,\,  7.5\pm  1.2$&$  1.9\pm  0.1$\\
      A133&0.056&01:02:41.68&-21:52:55.81&-84.2&$\,\,\,  5.5\pm  1.7$&$  1.7\pm  0.2$\\
      A780&0.055&09:18:05.67&-12:05:44.02&25.1&$\,\,\,  6.2\pm  2.5$&$  1.7\pm  0.2$\\
     A1781&0.062&13:44:52.56& 29:46:15.31&78.0&$\,\,\,  0.8\pm  0.5$&$  0.9\pm  0.2$\\
     A1795&0.063&13:48:52.58& 26:35:35.81&77.2&$\,\,\,  5.2\pm  1.0$&$  1.6\pm  0.1$\\
     A1991&0.059&14:54:31.50& 18:38:32.71&60.5&$\,\,\,  1.9\pm  0.5$&$  1.2\pm  0.1$\\
     MKW3S&0.044&15:21:51.85& 07:42:31.79&49.5&$\,\,\,  2.3\pm  0.6$&$  1.2\pm  0.1$\\
     \hline
\end{tabular}
%\end{adjustwidth}
\end{center}
\begin{list}{}{}
\item[$^{\mathrm{a}}$] Dynamical properties estimated by \citet{sifon15}.
\end{list}
\end{table*}

Each of the clusters is covered by a series of 2-minute exposures in the $g$ and $r$-bands. These exposures follow a large dither pattern and were taken with a cadence of typically a month to allow for a study of supernova rates in these clusters \citep{sand11,sand12,graham12}. This also ensures that background- and flatfield effects can be identified before producing the final stacks. The individual exposures are pre-processed using the \textit{Elixir} pipeline \citep{elixir04}. 20 frames are stacked in each band, resulting in a total exposure time of $\sim$40min. These frames were taken under conditions with very good seeing (typically all have PSF sizes FWHM$<$0.8$''$), as the data will be used to perform weak gravitational lensing measurements (C. S\'ifon et al., in prep.). Note that the analysis in this paper does not require data of this superb image quality, but it ensures that the relevant galaxies are easily resolved and thus allows for a study of their morphologies. We subtract large-scale residual background effects using an initial meshsize of 150 pixels ($\sim28''$), smoothed with a median filter of size 5$\times$5 meshes. This corrects background structures on a scale that is much larger than the sizes of the galaxies we study. For $g$-$r$ colour measurements we use a matched aperture approach in which we take account of the seeing differences between the two bands \citep[see Appendix A in][for details]{vdB13}. 

We do not have spectroscopic cluster membership information for UDGs in the cluster fields, so the CFHT Legacy Survey (CFHTLS) deep fields \citep[in particular the photometry from][]{erben09,hildebrandt09a} are used as a reference to perform a statistical background correction. By using the four independent fields we reduce and account for the effect of field-to-field variance in our analysis. By adding random pixel noise to the deep CFHTLS images, we mimic the depth of the cluster image stacks. We perform the same analysis on the CFHTLS depth-matched reference as on the cluster fields.

We automatically mask stars from the guide star catalogue 2.3 \citep{gsc} by placing circular masks at their locations, scaled in size according to the stellar magnitude. We also mask the predicted locations of their diffraction spikes by superimposing plus-shaped masks at the same locations, with length and width scaled with the magnitude. How the sizes of these masks scale with the stellar magnitude depends on the instrument, filter, and depth, and we determine this empirically for our setup.

After this automatic process, we manually mask primarily reflective haloes around the brightest stars, which have radii of about $3.5'$ for this instrument, and diffraction spikes of bright stars that fall outside of the MegaCam detector. During our analysis we take account of the reduced effective areas after masking. 

\begin{figure*}
\resizebox{\hsize}{!}{\includegraphics{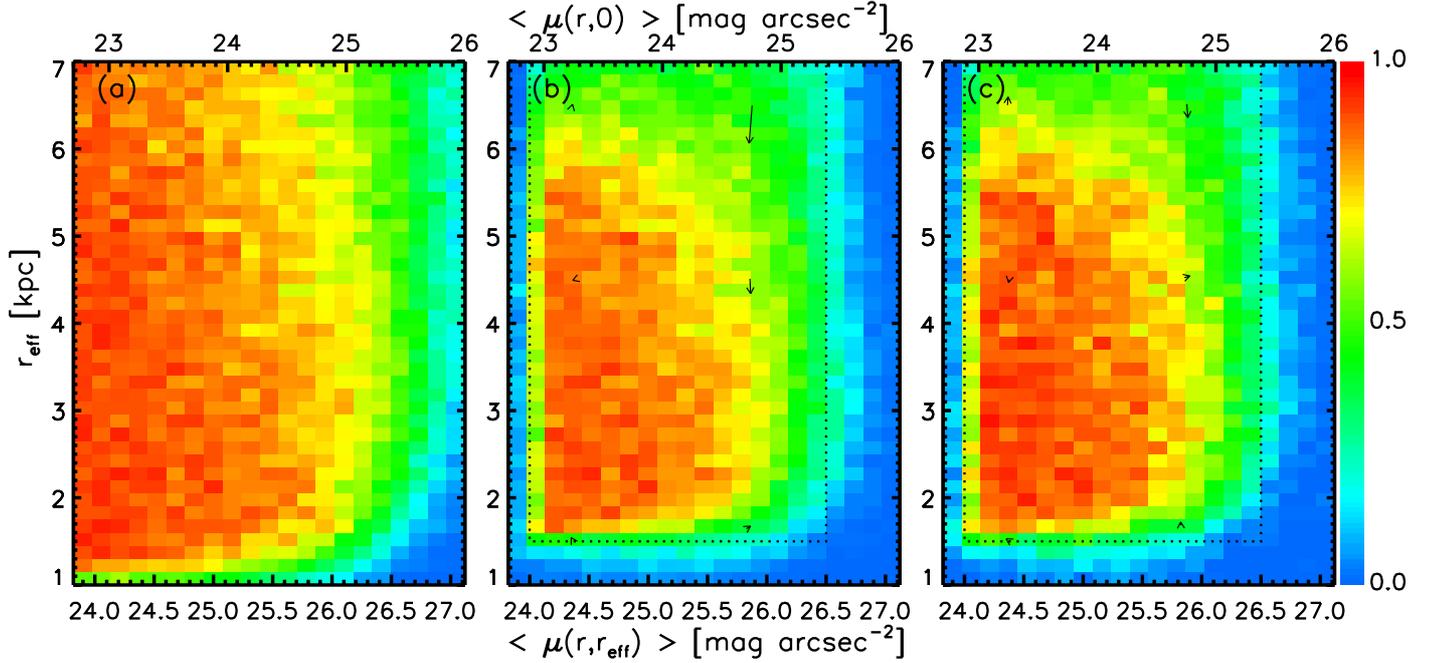}}
\caption{Recovered fraction of injected sources in the eight cluster fields, as a function of the circularized effective radius and the mean surface brightness within the effective radius. 80,000 sources have been simulated, with parameters drawn uniformly within the boundaries of the plot window, resulting in $\sim 150$ sources per histogram bin. \textit{(a):} Sources surviving the pre-selection based on \texttt{SExtractor} output parameters. \textit{(b)}: Same but after applying the selection criteria (listed in Table~\ref{tab:discarded}, and indicated by the dotted region) on the best-fitting parameters from \texttt{GALFIT}. Arrows show the (small) bias between intrinsic and recovered parameters in 6 parts of covered parameter space. \textit{(c)}: Same as panel (b), but here from the four CFHTLS Deep fields, with noise added to resemble the quality of the data for the cluster fields. Note that the distributions are not completely smooth due to obscuration from unmasked sources in the images, which prevents the detection of faint extended galaxies (slightly more so in the cluster fields). }
\label{fig:master3panel_new}
\end{figure*}

\section{Analysis}\label{sec:analysis}
We use \texttt{SExtractor} \citep{bertinarnouts96} on the $r$-band stack to detect sources. To improve our sensitivity towards faint extended galaxies, we use a Gaussian filter with a FWHM of 5 pixels ($\sim 1''$ with our detector), compared to the default one of 1.5 pixels. In order for a source to be ``detected'', we require it to have 20 pixels which are at least 0.70 sigma above the background (note that the noise is estimated before convolution with the filter). These values have been chosen to be as sensitive as possible to faint extended objects, while keeping the purity (of detecting real objects, and not noise fluctuations) high. The parameters were tuned on simulated images, with similar noise properties as our images. Using these detection parameters we expect $\sim 5$ noise fluctuations to end up in the \texttt{SExtractor} catalogues for each of the cluster fields. We find later that 95\% of these noise fluctuations do not pass the final selection criteria imposed after object detection, so these do not contribute in any significant way to the final sample we study. Although we would gain some depth by performing source detection on the combined $g+r$-band stack, this would make the selection somewhat colour-dependent, and for the interest of simplicity we leave that to future work. 

We detect $\sim 100,000$ sources per field (which is $\sim$20 arcmin$^{-2}$) using the optimised \texttt{SExtractor} configuration. The approach we follow is to proceed by (1) applying some simple criteria to the \texttt{SExtractor} output parameters to select against anything that is certainly not a UDG at the cluster redshift; (2) run \texttt{GALFIT} \citep{peng02} to estimate morphological parameters based on which the final selection of UDG candidates can be made. To keep a handle on the completeness of our UDG selection, we perform all processing steps also on a set of tailored image simulations.

\subsection{Image simulations}
Our main suite of image simulations is based on the injection of objects with S\'ersic profiles on random locations in the image stacks. We choose a constant S\'ersic index n=1, which is the baseline value measured for the UDGs found by \citet{vandokkum15} and \citet{koda15}. Sizes (effective radii, $\mathrm{r_{eff}}$) are drawn uniformly between 1 and 5 kpc, and central surface brightnesses uniformly in the range $23.0 < \mu(r,0)/[\mathrm{mag\, arcsec^{-2}}] < 26.0$. The ellipticity $\epsilon=(1-q)/(1+q)$, where $q=B/A$, is randomly drawn between 0 and 0.2, and we only consider simulated sources that have been injected on unmasked parts of the images. We inject only 2000 sources per simulation to ensure that these sources do not significantly affect the noise properties of the images, nor their own detectability. We repeat this experiment 10 times for each cluster, and perform 100 additional simulations of the central regions for each cluster to improve statistics. We perform additional simulations to test how sensitive our analysis is to the choice of e.g. the simulated S\'ersic index. 

Rather than working with central surface brightness, these additional simulations motivate us to work with a sligthly different quantity, namely $\langle\mu(r,\mathrm{r_{eff}})\rangle$, defined as the mean surface brightness \textit{within} the effective radius in the $r$-band. This parameter is always larger than the central surface brightness, and for a S\'ersic-index of 1 the difference is 1.12 mag arcsec$^{-2}$. In a plane of effective radius and mean effective surface brightness, each point then maps to a unique total magnitude, independently of its S\'ersic-index. Another advantage is that the $\langle\mu(r,\mathrm{r_{eff}})\rangle$ of a S\'ersic profile is more closely linked to the detectability of a source, in contrast to the \textit{central} surface brightness which increases rapidly for large S\'ersic-indices. We verify this by repeating the analysis of the recovery rate of simulated galaxies with S\'ersic index 0.5 and 1.5. We note further that the detectability of a source of a given effective (i.e. circularized) radius does not sensitively depend on its ellipticity. 

\subsection{Sample selection}\label{sec:sampleselection}
We apply a simple selection criterion based on a ratio of aperture fluxes measured with \texttt{SExtractor}. We select against stars and other compact objects by requiring that $r_{2''} > 0.9 + r_{7''}$, where $r_{x''}$ is the $r$-band magnitude within a circular diameter of $x$ arcsec. After applying the manual mask and this size criterion, we are left with $\sim 6000$ objects per field. We check the effect of these selection criteria on our simulated sources, see Fig.~\ref{fig:master3panel_new}a. This figure shows the recovery rate of simulated galaxies as a function of surface brightness $\langle\mu(r,\mathrm{r_{eff}})\rangle$ and circularized effective radius $\mathrm{r_{eff}}$. Incompleteness at low surface brightness is due to the limited depth of our data, and this is, at fixed surface brightness, most notable for smaller sources. Furthermore, the selection of small sources of any surface brightness has been suppressed by our aperture magnitude criterion. 

Without performing any by-eye inspection of selected candidates, we run \texttt{GALFIT} to find the S\'ersic profile that fits each galaxy best. We mask pixels belonging to any other detected object before the fitting procedure. The S\'ersic-n parameter is among the free parameters in the fit, and for injected profiles we find the latter to be recovered with a negligible bias when we leave the background as another free parameter. 

We now select galaxies with best-fit S\'ersic parameters in the range $24.0 \leq \langle\mu(r,\mathrm{r_{eff}})\rangle \leq 26.5 \, \mathrm{mag\, arcsec^{-2}}$, circularized effective radii $1.5\leq\mathrm{r_{eff}}\leq7.0$ kpc, that were fit within 7 pixels from the \texttt{SExtractor} position, and have a S\'ersic index n$\leq$4. We test these cuts on our simulated sources, which shows that the completeness within our region of interest remains high, see Fig.~\ref{fig:master3panel_new}b. Whereas almost all injected sources are recovered and correctly identified as UDGs, each real cluster image yields only $\sim 300$ UDG candidates out of the $\sim 6000$ \texttt{GALFIT} inputs. Note that this large drop is not surprising, given our pre-selection of large galaxies based on the \texttt{SExtractor} parameters. Most of the sources that are not selected as UDG candidates, are rejected because they have either too small sizes, or surface brightnesses that are typical for ``normal" galaxies at that redshift (cf. Table~\ref{tab:discarded}). As a test, we did run \texttt{GALFIT} on all $120\,000$ detected objects in the field of A85, and found that the same $\sim 300$ UDGs are eventually identified.

\begin{figure*}[!tbp]
  \centering
  \begin{minipage}[t]{0.31\textwidth}
    \includegraphics[width=\textwidth]{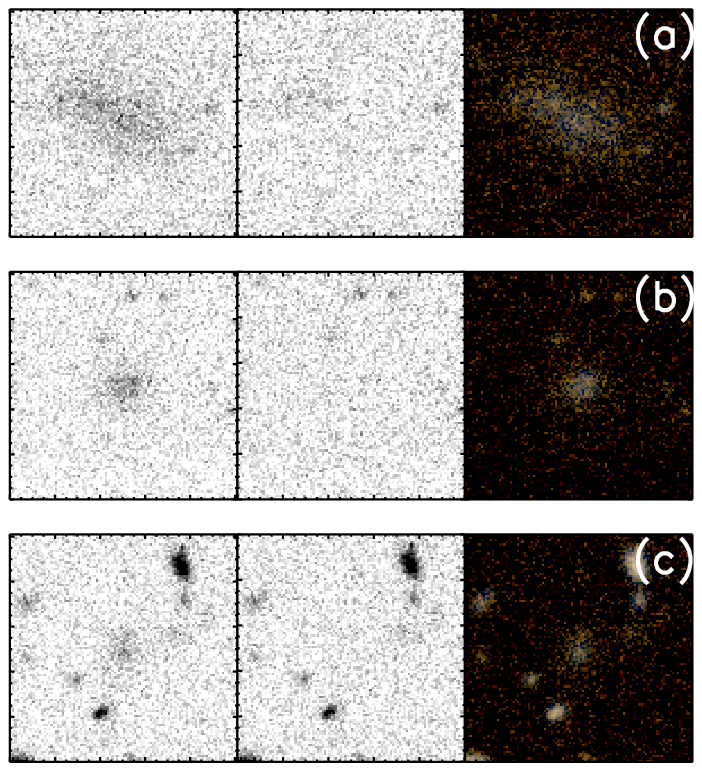}
    	    	  \end{minipage}\hfill
    	  	  \begin{minipage}[t]{0.31\textwidth}
    \includegraphics[width=\textwidth]{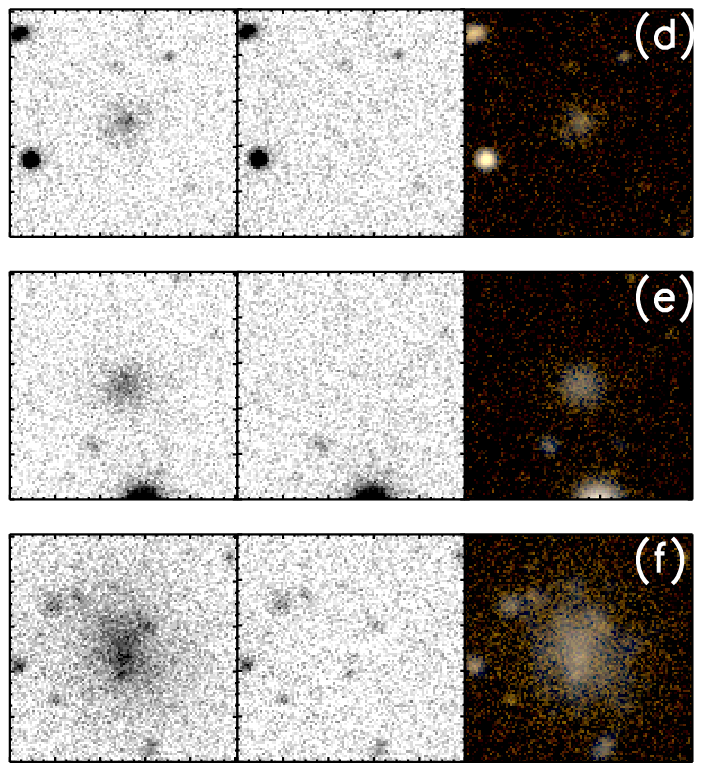}
  \end{minipage}\hfill
  \begin{minipage}[t]{0.31\textwidth}
    \includegraphics[width=\textwidth]{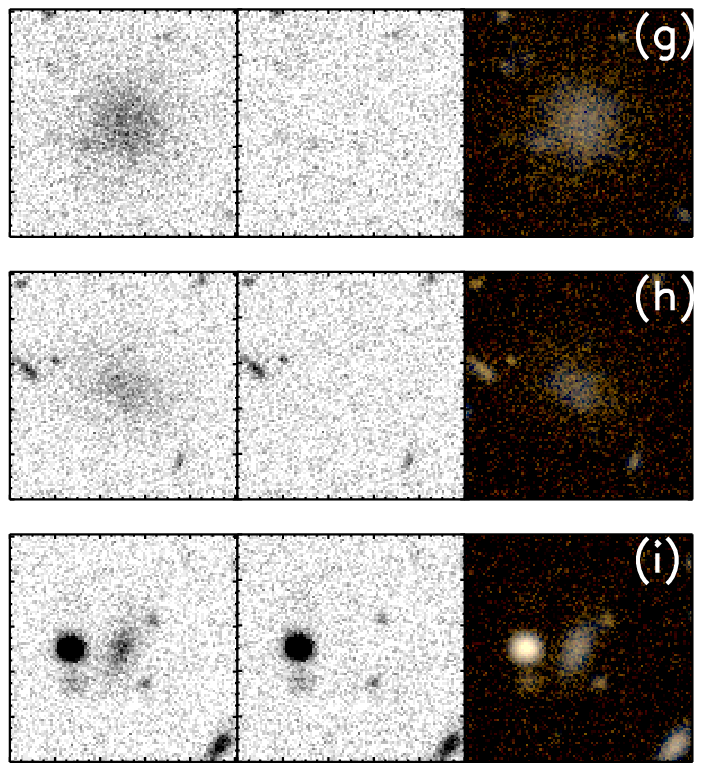}
  \end{minipage}
    \caption{Examples of nine typical galaxies we select. Images show 19$\times$19$''$ cutouts. \textit{Left panels:} processed original $r$-band images. \textit{Middle panels:} Residual of the $r$-band image after model subtraction. \textit{Right panels:} $g-r$ colour images. Best-fit morphological parameters and colours are listed in Table~\ref{tab:exampleparameters}.} 
  \label{fig:examples_paper}
\end{figure*}

The fact that the selection based on the \texttt{GALFIT} parameters has only a small effect on the simulated completeness (compare Figs.\ref{fig:master3panel_new}a~\&~\ref{fig:master3panel_new}b), yet greatly reduces the galaxies selected in the real data, indicates that the purity of this sample is remarkably increased. A quick by-eye scan of image cutouts confirms this. Given the pure sample of 2500 UDG candidates in the cluster fields, we decided not to perform any additional subjective selection criteria, so that our analysis remains reproducible. Figure~\ref{fig:examples_paper} shows several typical galaxies that satisfy our selection criteria. Note that the completeness in our selection of parameter space is similar between the noise-matched CFHTLS image and the clusters, see Fig.~\ref{fig:master3panel_new}c. In these matched CFHTLS fields there are $\sim$150 selected sources per field. 

\begin{table}%[ht]
\caption{Parameters that are used to select the sample of UDG candidates. Shown are the total numbers of rejected objects for the different selection criteria (combining all eight clusters).}
\label{tab:discarded}
\begin{center}
\begin{adjustwidth}{-0.4cm}{}
\begin{tabular}{l c c c}
\hline
\hline
Parameter & Requirement & Total discarded$^{\mathrm{a}}$ & Unique discarded$^{\mathrm{b}}$   \\
\hline
$\langle\mu(r,\mathrm{r_{eff}})\rangle$&  $\geq$24.0 $\mathrm{mag\, arcsec^{-2}}$&19328&4043\\
$\langle\mu(r,\mathrm{r_{eff}})\rangle$&  $\leq$26.5 $\mathrm{mag\, arcsec^{-2}}$&15399&604\\
$\mathrm{r_{eff}}$ &$\geq$1.5 kpc&26268&10007\\
$\mathrm{r_{eff}}$ &$\leq$7.0 kpc&8287&43\\
distSEx$^{\mathrm{c}}$&          $\leq$7.0 pix&932&122\\
S´\'ersic-n  &$\leq$4.0&19158&1186\\
\hline
\end{tabular}
\end{adjustwidth}
\end{center}
\begin{list}{}{}
\item[$^{\mathrm{a}}$] Total number of sources that do not meet this criterion.
\item[$^{\mathrm{b}}$] Number of sources that do not meet this criterion, but do meet all others. 
\item[$^{\mathrm{c}}$] Distance between best-fitting \texttt{GALFIT} position and the input position obtained using \texttt{SExtractor} (1 pix \^= 0.185$''$). If this difference is too large it is generally a sign that the \texttt{GALFIT} fit was unstable.
\end{list}
\end{table}

\begin{table}%[ht]
\caption{Best-fitting \texttt{GALFIT} parameters and $g-r$ colour of the examples shown in Fig.~\ref{fig:master3panel_new}.}
\label{tab:exampleparameters}
\begin{center}
\begin{tabular}{l c c c c}
\hline
\hline
&$\mathrm{r_{eff}}$&$\langle\mu(r,\mathrm{r_{eff}})\rangle$ & S\'ersic & $g-r$\\
& [kpc]  &  [mag arcsec$^{-2}$]  &n &  \\
\hline
(a)&4.28&25.75&0.76&0.46\\
(b)&2.47&25.62&1.19&0.50\\
(c)&1.56&25.67&1.15&0.48\\
(d)&1.52&25.18&0.89&0.83\\
(e)&1.72&25.13&0.96&0.50\\
(f)&5.42&25.32&1.22&0.61\\
(g)&2.96&25.22&0.72&0.59\\
(h)&3.33&25.62&0.76&0.65\\
(i)&2.26&25.06&1.33&0.75\\
\hline
\end{tabular}
\end{center}
\end{table}

In the next sections we study the properties of galaxies that satisfy our criteria to be ultra diffuse, and consider the contribution from galaxies in the fore- and background using the CFHTLS fields. We study several properties of the population in turn, which may help to provide clues where this population originates from and how it has evolved after being accreted onto the cluster.

\section{Colours and stellar masses}\label{sec:colours}
\begin{figure*}[!tbp]
  \centering
  \begin{minipage}[t]{0.49\textwidth}
    \includegraphics[width=\textwidth]{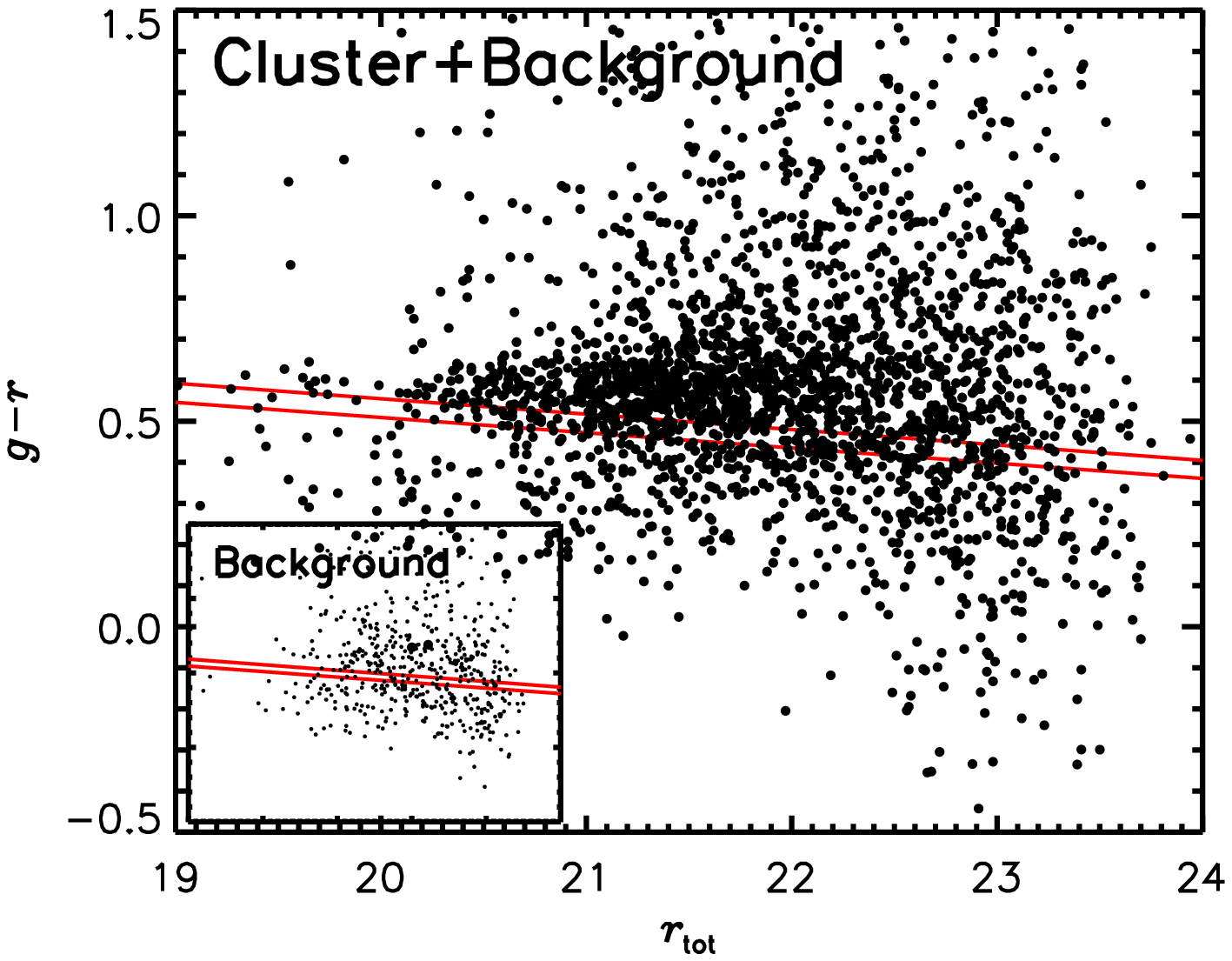}
    \caption{\textit{Main panel:} $g-r$ colour versus total $r$-band magnitude for the 2456 selected objects over 8 clusters. The red line marks the division line used in vdB15 to separate red-sequence galaxies from bluer galaxies for redshifts of $z=0.044$ and $z=0.063$, and should thus lie just below the red sequence. \textit{Inset panel:} The colour-magnitude distribution of sources selected in the same way from the depth-matched CFHTLS fields (same parameter range).}
    \label{fig:colourmag}
    	    	  \end{minipage}
    	  \hfill
  \begin{minipage}[t]{0.49\textwidth}
    \includegraphics[width=\textwidth]{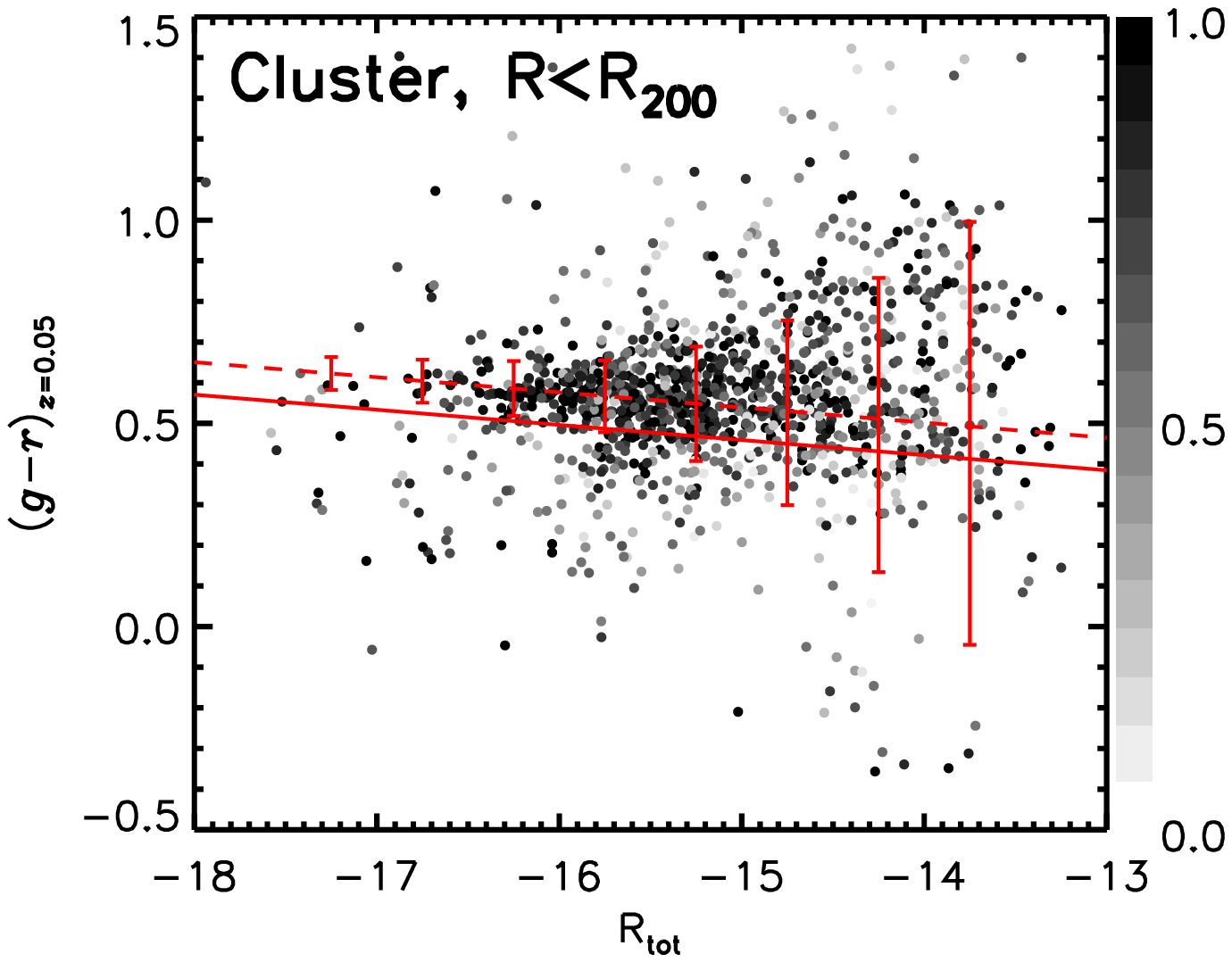}
    \caption{Same as Fig.~\ref{fig:colourmag}, but selecting sources within $R<R_{200}$, and subtracting the background statistically in each cluster field. Results are plotted as function of absolute $r$-band magnitude, and the ($g-r$)-colour is shifted to a common redshift of $z=0.05$. Statistical weights (after background subtraction) are indicated in grey scale. The red solid line is the division used in vdB15 to separate red-sequence from bluer galaxies (now independent of redshift due to the shift). The dashed line is the same but off-set by 0.08 to approximately fit the red sequence. Error bars show median 1-$\sigma$ uncertainties on the colour measurements, in bins of magnitude.}
    \label{fig:colourmag_abs}
  \end{minipage}
\end{figure*}
Colour measurements of selected UDGs help constrain the ages of their stellar populations, and are used to estimate stellar masses. Figure~\ref{fig:colourmag} shows the $g-r$ colour for this sample, plotted against the total \texttt{GALFIT}-integrated $r$-band magnitude. The red lines are used in vdB15 to separate red-sequence galaxies from those that are bluer, for reference redshifts of $z=0.044$ and $z=0.063$. Even though the sample of galaxies we consider is selected based only on their size versus surface brightness distribution in the $r$-band, there is a clear overdensity of sources in this diagram, just above the red lines. This shows that they lie on the extension of the red sequence, as it was identified for brighter galaxies in vdB15. It is worth noting that the galaxies that are shown here are typically a factor hundred fainter than galaxies around the characteristic magnitude, which is around $15<r_\mathrm{tot}^{*}<16$ at these redshifts \citep[e.g.][]{blanton03}. 

Also shown is a small inset figure with the colour distribution of galaxies satisfying the same morphological criteria in the CFHTLS fields, matched in depth. Note that the main panel is not yet corrected for the presence of such background galaxies. The CFHTLS fields show a more widely spread colour distribution without a clear red sequence, because the sources are not expected to be at a single redshift. We now consider all UDGs in the cluster fields, with projected radii R$<$R$_{200}$, where estimates of R$_{200}$ are based on a dynamical analysis of galaxy redshifts \citep{sifon15}, and are presented in Table~\ref{tab:overview}. The field background is subtracted point-by-point statistically from the cluster distribution, as follows. For each cluster we measure the unmasked area within R$_{200}$. We determine a weight for each background point by comparing the total area of the CFHTLS fields to this unmasked area within R$_{200}$. After setting the weight of each cluster point to 1.0, we take each source in the background fields, and subtract its weight from the nearest cluster point in this colour-magnitude diagram. Once a point has reached a weight=0, we find the second closest neighbour, etcetera. Figure~\ref{fig:colourmag_abs} shows the distribution of cluster points, after this background subtraction. The $x$-axis now shows the absolute magnitude in the $r$-band, while the $y$-axis shows the ($g$-$r$)-colour, shifted in colour to a redshift of $z$=0.05\footnote{Note that this is similar to, but a simplified version of, a K+E-correction. Because we know the colour of the red sequence at different redshifts, we simply attribute the evolution to a $g-r$-colour shift and correct for this, while a standard K+E-correction takes part of the correction as being along the $x$-direction because of a small passive evolution of the total absolute magnitude. We note that this simplification has no significant effect on the analysis.}. The red solid line is the division used in vdB15 to separate red-sequence from bluer galaxies (now independent of redshift due to the correction). The dashed line is the same but off-set by 0.08 to approximately fit the red sequence. Error bars show median 1-$\sigma$ uncertainties on the colour measurements, in bins of magnitude. 

\begin{figure}
\resizebox{\hsize}{!}{\includegraphics{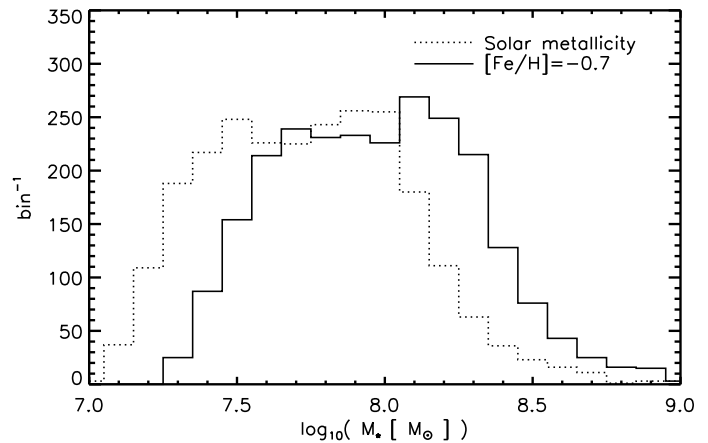}}
\caption{Estimated stellar mass distribution for the selected sample of 2456 galaxies, assuming all galaxies are at the redshift of the clusters. A single stellar mass-to-light ratio is assumed, corresponding to a dust-free stellar population with colour $g-r$=0.6, for the \citet{chabrier03} IMF, and two different metallicities.}
\label{fig:stelmassdist}
\end{figure}

The $g-r$ colours indicate that these galaxies almost all lie on the red sequence\footnote{However note that some of the selected galaxies have colours off-set from the red sequence, such as the bright, blue galaxies around $\mathrm{R_{tot}\simeq -17}$ and $(g-r)_{z=0.05}\simeq 0.2$. We inspected these by eye, and they appear to be blue galaxies in the process of merging, which star-forming activity may be merger-induced.}, and thus likely contain relatively old stellar populations. Note that there is a degeneracy between stellar age and metallicity \citep[e.g.][]{worthey94}, which we cannot break with these photometric data. Assuming solar metallicity and no dust, the \citet{bc03} models require a stellar population with an age of 2 Gyr to reproduce a colour of $g-r=0.6$. Low-mass galaxies typically have lower stellar metallicities \citep[e.g.][]{kirby13}. As an example, if we assume sub-solar metallicity of \hbox{[Fe/H]=-0.7}, and no dust, the age of the stellar population is around 6 Gyr. These assumptions also affect the stellar mass estimates, but to a lesser extent. Figure~\ref{fig:stelmassdist} shows the stellar mass distribution of the UDGs, under the assumption that all galaxies are at the redshift of the cluster. The two distributions correspond to the different assumed metallicities (and corresponding ages), and differ by a factor $\sim$1.7. Assuming a metallicity of \hbox{[Fe/H]=-0.7}, the median stellar mass of the selected galaxies is $\sim 10^{8}\,\mathrm{M_{\odot}}$.

We stress that the studied sample is defined by a selection in effective size and surface brightness. Because of the large size range considered ($1.5\leq\mathrm{r_{eff}}\leq7.0$ kpc), we naturally end up with a distribution that quite gradually falls off both at high and low masses. The selected objects with the highest masses are simply the large galaxies with high surface brightnesses (cf. e.g. upper left corners of Fig.~\ref{fig:master3panel_new}), and the lowest-mass objects have small sizes and low surface brightnesses. To be able to define a sensible completeness, in terms of mass in Fig.~\ref{fig:stelmassdist}, or in terms of magnitude in Figs.~\ref{fig:colourmag}~\&~\ref{fig:colourmag_abs}, knowledge of the underlying size and surface brightness distributions is thus required. This is the objective of Sect.~\ref{sec:sizedist}.

\section{Abundance and size distribution}\label{sec:abundancehalomass}
Our best information so far has been limited to the study of UDGs in the Coma cluster (and to some extent the Virgo cluster). Given our cluster sample of eight, we can study for the first time the abundance of UDGs as a function of cluster properties. Table~\ref{tab:abundances} gives the measured number of UDGs, satisfying our selection criteria, within a projected radius of R$_{200}$ for each of the eight clusters. When we again perform a statistical background subtraction, we obtain the numbers in the column ``Corrected''. These are the same number of UDGs as shown in Fig.~\ref{fig:colourmag_abs} (taking the weights into account). Within these projected radii, 30-190 UDGs per cluster are detected. These clusters are at different (low) redshifts, but note that this only has a small effect on the detection limits. The reason is that the surface brightness limits of the different clusters are comparable (and independent of redshift), and the angular diameter and luminosity distances increase still roughly 1:1 at such low redshifts. The latter means that a galaxy with given intrinsic properties retains its effective surface brightness when moved to higher (but still low) redshift. In practice this breaks down for higher redshifts, when the PSF size becomes substantial compared to the intrinsic size of the galaxies. Given that the effect is small, we do not attempt to correct for this in the current study.

\begin{figure}
\resizebox{\hsize}{!}{\includegraphics{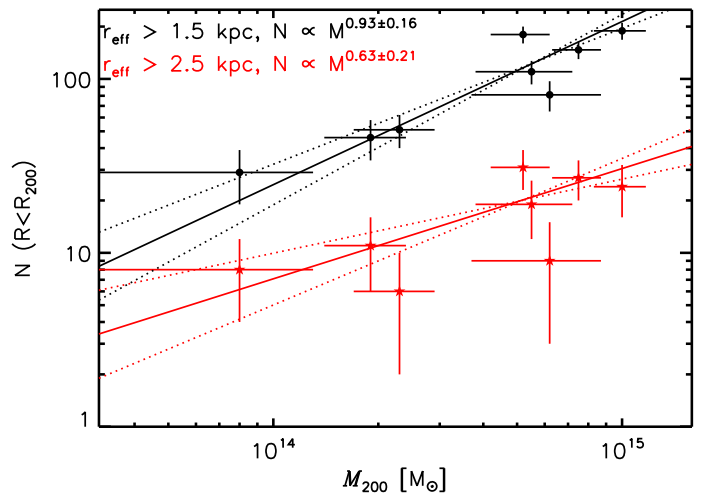}}
\caption{Abundance of UDGs as a function of halo mass, for two different size cuts. \textit{Solid lines:} best-fitting power-law relations between number density and halo mass, for each size cut. \textit{Dotted lines:} relations corresponding to the $\pm 1$-$\sigma$ uncertainties on the power-law index.}
\label{fig:abundancehalomass}
\end{figure}

Figure~\ref{fig:abundancehalomass} shows the abundance of UDGs, as a function of halo mass (cf. Tables~\ref{tab:overview}~\&~\ref{tab:abundances}). This shows a clear correlation between the abundance of UDGs and their host halo masses, with a best-fit power-law relation of $N\propto M^{0.93\pm0.16}$. When increasing the size cut, we find that the overall abundance decreases, but the slope remains similar within the uncertainties. So even the largest, most diffuse galaxies seem to be abundant in each of the eight clusters. The abundance drops rapidly when we consider larger and larger galaxies at these surface brightnesses. Next, we quantify the underlying size distribution. 

\begin{table}%[ht]
\caption{Number of UDGs with projected R$<$R$_{200}$. Each cluster shows a significant overdensity of the galaxies considered.}
\label{tab:abundances}
\begin{center}
\begin{tabular}{l c c }
\hline
\hline
Cluster & Raw counts$^{\mathrm{a}}$ & Corrected$^{\mathrm{b}}$   \\
\hline
A85 &          313 &$          189 \pm           21 $\\
A119 &          221 &$          147 \pm           17 $\\
A133 &          207 &$          110 \pm           17 $\\
A780 &          173 &$   \,\,\,        81 \pm           16 $\\
A1781 &           62 &$  \,\,\,         29 \pm           10 $\\
A1795 &          288 &$          180 \pm           20 $\\
A1991 &           97 &$     \,\,\,      46 \pm           12 $\\
MKW3S &           85 &$       \,\,\,    51 \pm           11 $\\
\hline
\end{tabular}
\end{center}
\begin{list}{}{}
\item[$^{\mathrm{a}}$] Raw numbers, neither corrected for incompleteness, nor background subtracted.
\item[$^{\mathrm{b}}$] After subtracting the background from the CFHTLS fields, matched in depth, for the same redshift and matched in unmasked area. Poisson errors, which are given, dominate over the field-to-field variance that we measure from the four spatially-independent CFHTLS fields.
\end{list}
\end{table}

\subsection{Size distribution}\label{sec:sizedist}
The discovery of large and diffuse galaxies in the Coma and Virgo clusters leads us to question whether they are part of a continuous size distribution, or whether they are a population that is isolated in parameter space (size versus surface brightness). Figure~\ref{fig:abundancehalomass}, in which the abundance for different size-cuts is shown, starts to address this question, but the way our sample is constructed allows a more quantitative approach.

\begin{figure}
\resizebox{\hsize}{!}{\includegraphics{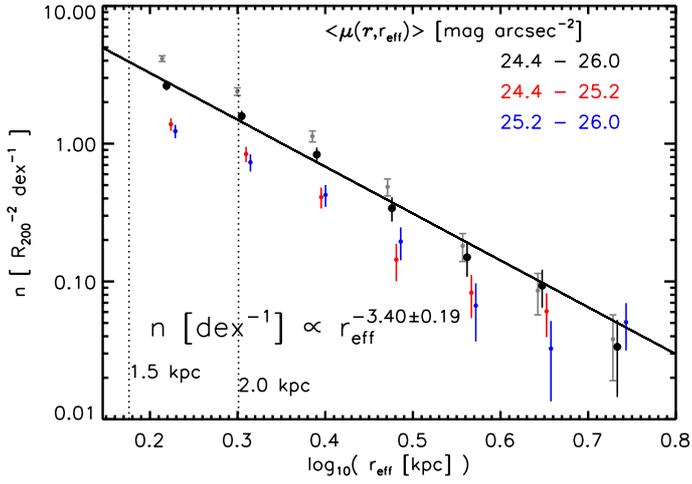}}
\caption{Measured size distribution of galaxies in the central projected $R_{200}$ for the studied clusters, before (grey) and after (black) statistical field subtraction. \textit{Blue and red points:} measured sizes for galaxies split over two bins of mean effective surface brightness, also after correction. In this size-range, the data are well described by a power-law index of -3.40$\pm$0.19 (in bins of equal logarithmic size).}
\label{fig:sizedist}
\end{figure}

The number density of the selected galaxies in the ensemble cluster with R$<$R$_{200}$, as a function of their effective radii, is shown by the grey points in Fig.~\ref{fig:sizedist}. In this figure, we restrict ourselves to surface brightnesses of $24.4 < \langle\mu(r,\mathrm{r_{eff}})\rangle < 26.0$ mag arcsec$^{-2}$, since we are highly complete in this region of parameter space (cf. Fig.~\ref{fig:master3panel_new}b). The figure suggests that there is a continuous distribution of galaxies in the size-surface brightness plane, and shows that the number density of UDGs in a certain size bin is always dominated by the smaller galaxies in the bin. 

The black points are the same data, after subtracting the background statistically, and thus correspond to the cluster population. Although there is only a small bias between measured and intrinsic morphological parameters, measurement uncertainties may have a significant effect on the observed size distribution (\citet{eddington1913} bias, which is expected due to the steep slope). Therefore, the black points also include a small correction for this bias, which is at most 40\% per data point, and is estimated in the following way. We assume a size distribution of the form $\mathrm{n\propto r_{eff}^{\alpha}}$, as motivated by the distribution of the grey points, and give our simulated galaxies weights to emulate this size distribution. We then perform the complete selection process, as described in Sect.~\ref{sec:sampleselection}, on these simulated galaxies. The fractional difference between the intrinsic and the retrieved size distribution of these simulated galaxies is then used to correct the measured data points. We keep adjusting the assumed intrinsic slope $\alpha$, correct and fit the data points until the slope converges to $\alpha=-3.40\pm0.19$ with a $\mathrm{\chi^2/d.o.f.}=1.12$. Note that without correction we would have obtained a similar slope of $\alpha=-3.33\pm0.22$, but with a larger $\mathrm{\chi^2/d.o.f.}=1.77$. The steep slope that we find is in qualitative agreement with the size distribution presented in Fig.~4d of \citet[][]{koda15}, but note that their distribution is not corrected for incompleteness. However, it differs from the UDGs distribution in \citet[Figure 3 of][]{vandokkum15}, which may be due to their (partly subjective) selection criteria.

The same analysis is performed on two subsamples with mean effective surface brightnesses of $24.4 < \langle\mu(r,\mathrm{r_{eff}})\rangle \leq 25.2$, and $25.2 < \langle\mu(r,\mathrm{r_{eff}})\rangle < 26.0$, indicated by the red and blue points in Fig.~\ref{fig:sizedist}. The size distributions of these two subsamples follow the same logarithmic slope, within uncertainties, as the main sample. Moreover, the blue and red points have a similar normalisation, both containing about 50\% of the galaxies of the parent sample. Interestingly, this shows that, at a given effective radius for the sample we study, the distribution seems to be uniform as a function of mean effective surface brightness. This seems to be qualitatively consistent with Figure 4d of \citet{koda15}, although incompleteness in their diagram prevents a direct comparison.

In conclusion, in the range of parameter space (size and effective surface brightness) we consider, the UDGs seem to be part of a continuous distribution. The largest galaxies only constitute a small fraction of the total population, highlighting the rarity of the larger UDGs.

\section{Radial distribution}\label{sec:radialdist}
\begin{figure}
\resizebox{\hsize}{!}{\includegraphics{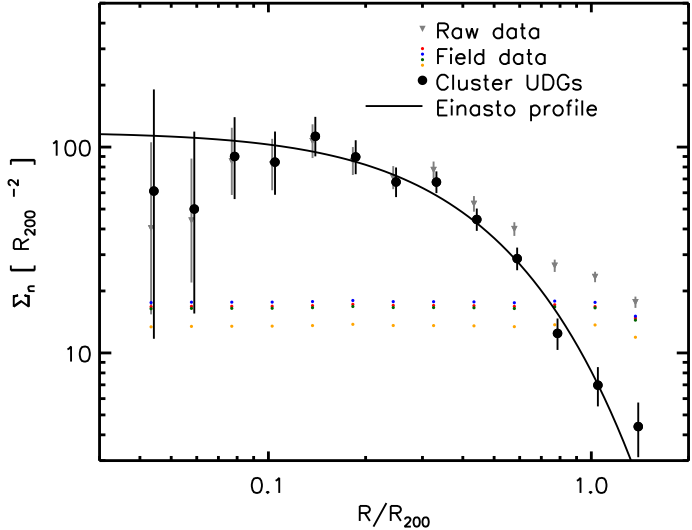}}
\caption{\textit{Small grey triangles:} Radial number density distribution of the selected sample of UDGs for the ensemble cluster field. \textit{Four colours:} Reference depth-matched CFHTLS fields. \textit{Large black symbols:} Background-subtracted and completeness-corrected UDGs in the ensemble cluster. \textit{Solid curve:} Best-fit projected Einasto profile to the cluster data.}
\label{fig:radialplot}
\end{figure}

The distribution of UDGs as a function of clustercentric distance may provide clues to their formation mechanism, and primarily their evolution and eventual fate in the cluster environment. \citet{vandokkum15} did not detect any UDGs within a projected distance of 300 kpc from the centre of Coma, but cautioned that this may be due to the images being shallower in the centre due to confusion and the extended light profiles from bright galaxies and the Intracluster Light (ICL). Nonetheless, this limit was used as input in the numerical simulation by \citet[][]{yozinbekki15}, and was also used to put an approximate constraint on the binding mass of UDGs in \citet{vandokkum15}. Because of its potential to be used as an input, and a test, to numerical simulations, it is important to measure the radial distribution of UDGs while taking account of the reduced completeness in the cluster centres. 

The radial number density distribution of selected UDGs in the sample of eight clusters is shown in Fig.~\ref{fig:radialplot}. The grey triangles are the raw numbers (i.e. neither corrected for the presence of background galaxies, nor for incompleteness) measured in each bin, normalised by their unmasked areas, and centred on the locations of the BCGs. The smaller coloured points give the mean number densities in the depth-matched CFHTLS fields\footnote{Note that these are not exactly constant per bin, because R$_{200}$ corresponds to a different angular size for each cluster, and therefore the masked area has a different effect on the background counts. Also for larger radii the background goes down because the lower redshift clusters do not have optical data that extend far enough, and the background counts drop for clusters at higher redshifts.}. The distribution of UDGs shows a flattening near the cluster centres. However, in contrast to \citet{vandokkum15}, we do find UDGs within a projected distance of 0.15$\times R_{200}$ (which corresponds to roughly 300 kpc), in numbers that are significantly higher than the background counts. %Part of the observed central deficit may be explained by incompleteness, and 
In Appendix~\ref{sec:radcompletenesscorr} we describe in detail how we correct the data for incompleteness.

Starting from the raw UDG number counts in the cluster fields, we subtract the background from the CFHTLS and correct the counts for the effective selection efficiency, as explained in Appendix~\ref{sec:radcompletenesscorr}. The resulting numbers are the cluster UDGs, shown with the large black circles in Fig.~\ref{fig:radialplot}. The projected distribution is shallow at small radii (R$<$0.2$\times R_{200}$), and falls off steeply at larger radii (R$\gtrsim$0.6$\times R_{200}$). None of these effects are significantly affected by our completeness correction, or the assumed size distribution. The behaviour is not well described by a projected NFW \citep{NFW} profile, and we do not find a good fit using this parameterisation. Given its extra degree of freedom, an \citet{einasto65} profile does provide a good description of the data. The best-fitting parameters are $c_\mathrm{EIN}=1.83^{+0.13}_{-0.12}$ and $\alpha_\mathrm{EIN}=0.92^{+0.08}_{-0.18}$, with $\mathrm{\chi^2/d.o.f.}=0.88$. Although this profile provides a good description of the data, we note that the best-fitting parameters are radically different from, and in particular the curvature parameter $\alpha$ is much larger than, the Einasto profiles that are used to describe the total, or dark matter, distribution of galaxy clusters \citep{dutton14,klypin16}.

\begin{figure*}[!tb]
  \centering
  \begin{minipage}[t]{0.49\textwidth}
    \includegraphics[width=\textwidth]{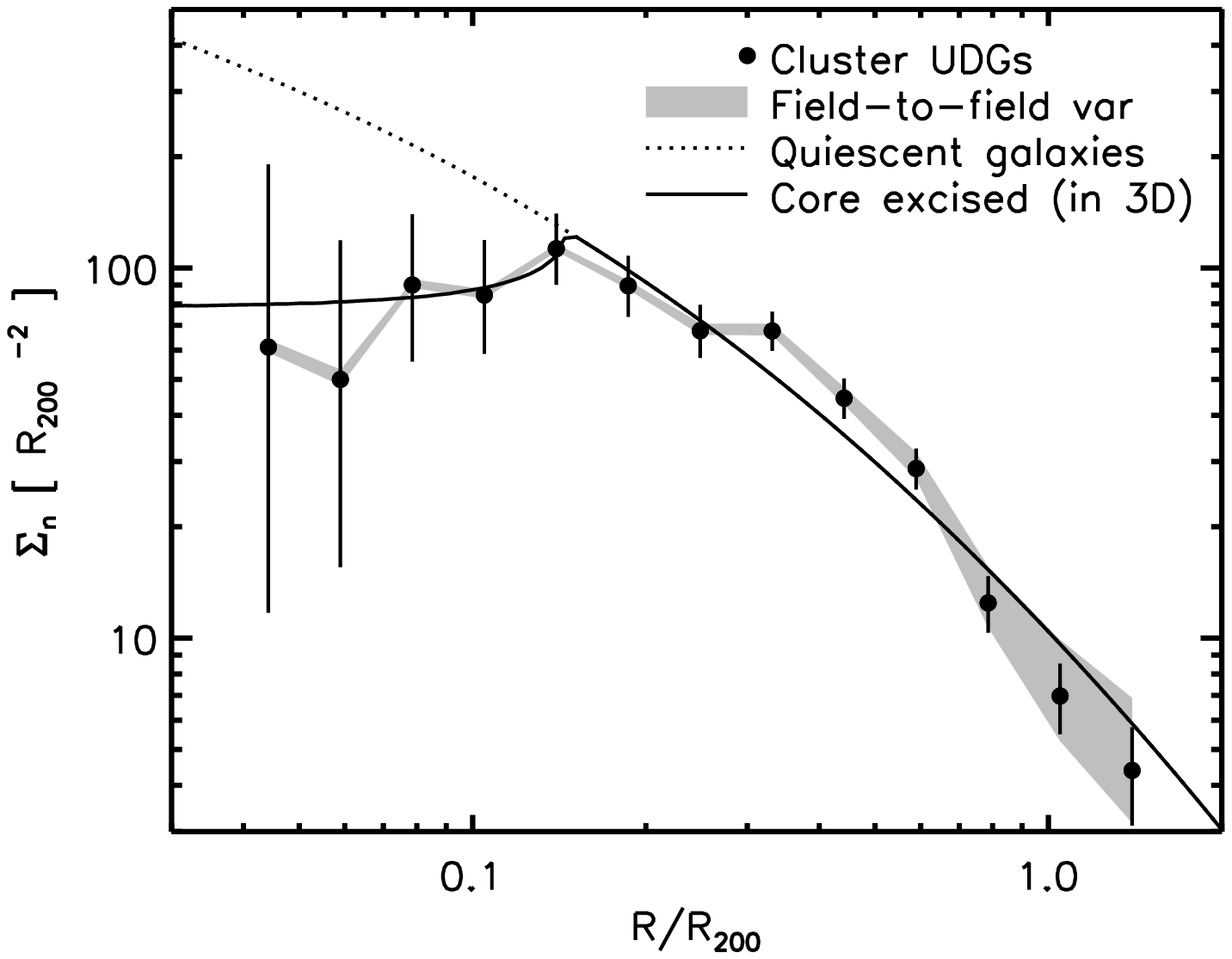}
\caption{\textit{Black points with statistical errors:} Background-subtracted and completeness- corrected UDGs in the ensemble cluster (as in Fig.~\ref{fig:radialplot}). \textit{Grey area:} the systematic error due to field-to-field variance in the reference background fields. \textit{Dotted curve:} Best-fit projected generalized NFW profile to the stellar mass distribution of the total population of quiescent galaxies in the clusters (vdB15), rescaled in normalisation. \textit{Black curve:} Same but with an excised core of 0.15$\times R_{200}$ (before projection).}
    \label{fig:radialplot2}
  \end{minipage}
  \hfill
    \begin{minipage}[t]{0.49\textwidth}
    \includegraphics[width=\textwidth]{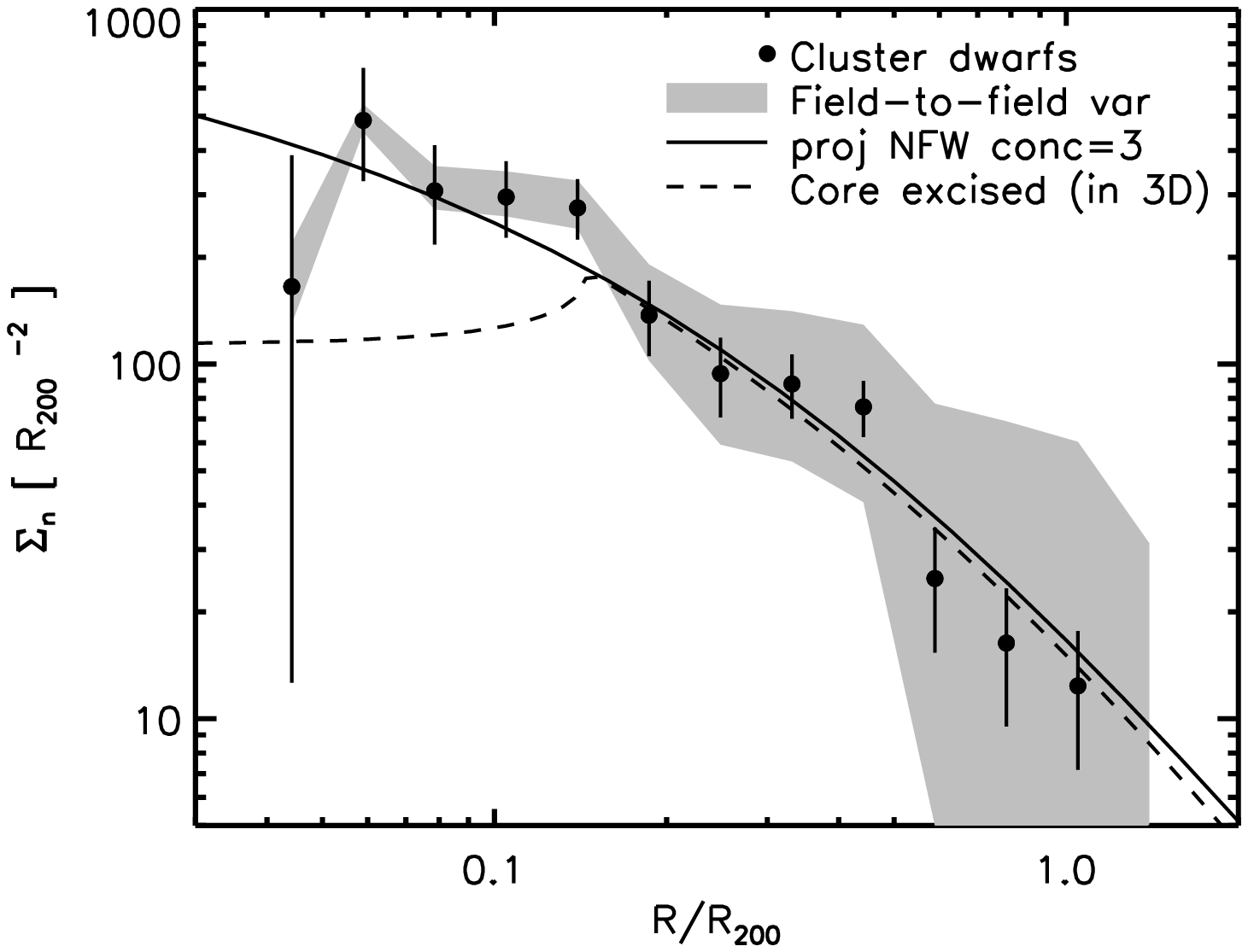}
\caption{Same as Fig.~\ref{fig:radialplot2}, but for regular dwarf galaxies with similar luminosities as the UDGs. \textit{Dashed curve:} Same core-excised profile of Fig.~\ref{fig:radialplot2}, scaled in normalisation (by a factor 1.45 compared to the UDGs) to best fit the data outside the core. \textit{Solid curve:} Projected NFW profile with concentration parameter $c_\mathrm{NFW}=3$, scaled in normalisation.}    \label{fig:radialplot_dwarfs}
    	    	  \end{minipage}	
\end{figure*}

\subsection{Comparison to more massive galaxies}
Though the UDGs do not seem to trace the overall matter distribution, comparing the estimated number density distribution of UDGs to the radial distribution of other galaxies in the clusters may provide insights into their origin and evolution. For instance, vdB15 compare the radial distribution of red-sequence galaxies to the distribution of bluer galaxies. Whereas red-sequence galaxies completely dominate the galaxy distribution in the centres, the distribution of bluer galaxies is less concentrated. This is related to their relatively recent accretion onto the cluster, and the observational fact that their quenching time is much shorter than the dynamical friction timescale. The radial distribution of a galaxy population thus provides information on the time of accretion, but also contains information on dynamical interactions happening inside the clusters.

The UDGs follow a radial distribution that is not well described by either of the distributions studied in vdB15. The outer part of the UDG radial profile is significantly steeper than the projected distribution of star-forming galaxies. On the other hand, they seem to follow a similarly steep slope as the full population of red-sequence galaxies, but there is a significant central deficit compared to this distribution. 

We consider a scenario in which the UDGs follow the dynamically old population of red-sequence galaxies studied in vdB15, but suffer from a depletion in the centre. Physical explanations for such a depletion may be the strong dynamical interactions between the dense galaxy population in the cluster centres and/or their tidal destruction caused by the overall cluster potential. As a simple model, we thus take the red curve of Fig.~8 in vdB15, which is described by a generalised NFW profile with parameters $c_\mathrm{gNFW}=2.39$ and $\alpha_\mathrm{gNFW}=1.31$, and which follows the full stellar-mass-density-weighted distribution of quiescent galaxies in these clusters. To model the central depletion, we set the density within a radius $\mathrm{r_{core}}$ to zero. Note that this radius is defined in 3D, after which the profile is projected along the line-of-sight. We thus fit only two free parameters, being the overall normalisation, and $\mathrm{r_{core}}$. The fit to the data is reasonably good, with $\mathrm{\chi^2/d.o.f.}=1.42$, for $\mathrm{r_{core}}=0.15^{+0.01}_{-0.02}\times R_{200}$. This corresponds to a physical radius of $\sim$300 kpc.

Note that we re-normalised the best-fit relation to the total stellar mass distribution in quiescent galaxies in these clusters (from vdB15). Based on the re-scaling we find that there is 1 UDG per $5\times 10^{10}\,\mathrm{M_{\odot}}$ of stellar mass in quiescent galaxies. Given the median mass of the UDGs in our sample, and the dominance of quiescent galaxies in these low-$z$ clusters (in terms of stellar mass), we find that the UDGs we select comprise about 0.2\% of the total stellar mass of the clusters.

\subsection{Comparison to normal dwarfs}\label{sec:comparisondwarfs}
The results presented above lead us to wonder how the radial distribution of UDGs compares to that of similarly luminous but more compact galaxies. In this Section we thus present a separate analysis to provide this comparison. We do no longer perform a pre-selection of dwarf candidates using \texttt{SExtractor}, but instead run \texttt{GALFIT} on every source detected on the cluster-, and reference depth-matched field-, images. We select sources with total integrated magnitudes in the range $21.0\leq r_\mathrm{tot} \leq 22.5$ (i.e. total magnitudes that are typical for the majority of our selected UDGs, cf. Fig.~\ref{fig:colourmag}), but with smaller effective radii of $0.5\leq\mathrm{r_{eff}}\leq1.0$ kpc \citep[i.e. more typical sizes for dwarf galaxies, see e.g. the compilation of][]{misgeld11}. After verifying that the $g-r$ red sequence has the same colour as for the UDGs, we apply a colour cut of 0.3$<g-r<$0.8 to improve the contrast of cluster dwarfs against the background. The large black data points in Fig.~\ref{fig:radialplot_dwarfs} represent the background-corrected dwarf galaxy distribution in the same radial bins as before. Error bars show Poissonian uncertainties, but the dominant uncertainty caused by field-to-field variance is indicated by the grey shaded area. The contrast with respect to the background of normal dwarf galaxies is much lower than that of the larger UDGs, even after applying the colour selection. That is because normal background galaxies at higher redshift appear fainter and smaller, and thus contaminate the selection of the cluster dwarf galaxies.

As before, the black data points also include a radially-dependent completeness correction, which we base on the injection and recovery of simulated dwarf galaxies with similar S\'ersic profile parameters as the ones we study here. The estimated completeness at the cluster outskirts is $\sim$85\%, and is thus notably higher than for the more extended galaxies with the same total magnitude. The completeness drops towards $\sim$60\% in the central bin, mostly due to obscuration by larger cluster galaxies. We performed two sets of simulations with S\'ersic-n=1.0 and S\'ersic-n=4.0 to verify that the result is robust with respect to the assumed parameter, which it is.

The dashed line shows the same core-excised model as in Fig.~\ref{fig:radialplot2}, but scaled up by a factor of 1.45 compared to the UDGs to best fit the dwarf galaxy data at $R>0.15\times R_{200}$. This moderate factor indicates that the UDGs are relatively abundant compared to these more compact dwarfs. If we would extend the power law fitted to the size distribution of UDGs in Fig.~\ref{fig:sizedist} down to $\mathrm{r_{eff}<1.0\,kpc}$, we would expect $\sim 15$ times more compact dwarfs in this size and magnitude range than we measure. We leave a full investigation of the size distribution that spans the entire size range to a future study.

The radial distribution of the selected dwarf galaxies does not show the same dramatic reduction in number density in the core as the more diffuse galaxies of similar luminosities. The model that describes the UDGs well, does not describe the regular dwarf galaxies in the core. We also fit a projected NFW model, with fixed logarithmic inner slope of -1, with fixed concentration of $c=3$ \citep[i.e. roughly what is expected for relaxed haloes of these masses in the local universe, e.g.][]{duffy08}, as shown by the black solid line. Although the uncertainties, especially due to field-to-field variance, are large, we find such a model to describe the data well ($\mathrm{\chi^2/d.o.f.}=1.07$). In contrast to the UDGs, the more compact dwarf galaxies therefore seem to trace the overall dark matter distribution.

\section{Discussion}\label{sec:discussion}
The abundance of UDGs is significant in each of the eight clusters we study, which argues that they are a common phenomenon in galaxy clusters. The question remains in which environments these galaxies have formed. In this respect, it is interesting to examine whether the trend with halo mass that we find down to $M_{200} \simeq 10^{14}\,\mathrm{M_{\odot}}$ (cf. Fig.~\ref{fig:abundancehalomass}) extends to even lower mass haloes. Whether UDGs also exist as isolated galaxies in the field is yet unclear. Although the reference fields contain galaxies with similar observed properties, redshift measurements are required to determine their physical sizes, and to better probe their local environment. 

The clusters we study provide an environment that allows us to learn more about their properties after being accreted onto a massive halo. The ICL of massive haloes has been observed to extend roughly up to a radius of 100-400 kpc \citep[][although no sharp edge is visible]{mihos05,gonzalez05,presotto14,demaio15}, and this is expected to be the result of the tidal disruption and stripping of stellar material from infalling galaxies. The cluster-centric distance at which this happens, for a given satellite, is related to its binding energy \citep{contini14}. We find the UDGs to trace the overall galaxy populations down to a cluster-centric distance of $\mathrm{r} \simeq 0.15 \times R_{200}$ (or $\sim$300 kpc) from the cluster centres, and to be consistent with zero within this region. Assuming that this means that they are severely tidally disrupted and eventually dissociated within this region, this indicates that these UDGs provide a small contribution to the ICL content. 

We estimate the mass required in these galaxies to withstand tidal disruption by the cluster down to this radius of $D\simeq 300$ kpc, following similar arguments as presented in \citet{vandokkum15}. The minimum mass required within a radius $r_{\mathrm{disrupt}}$ to resist a tidal field from a total mass $M$ at a radial distance $D$ from the cluster centre can be approximated by \citep{binney87}
\begin{equation}
\label{eqn:tidalmass}
m \gtrsim 3M \left(\frac{r_{\mathrm{disrupt}}}{D}\right)^3.
\end{equation}
The galaxies have smooth morphologies up to at least two times their stellar effective radii, so we consider $r_{\mathrm{disrupt}}=6$ kpc. At distances larger than $D\gtrsim 100$ kpc, we ignore the stellar mass of the BCG (which is about $M_{\star}\simeq 10^{12}\,\mathrm{M_{\odot}}$, see vdB15) compared to the interior mass expected for an NFW \citep{NFW} profile with a total mass of  $M_{200}\simeq 5\times 10^{14}\, \mathrm{M_{\odot}}$ (i.e. the median halo mass of our sample, cf. Table~\ref{tab:overview}), which is expected to be at least 10 times larger at these distances. This means that the mass required within $r_{\mathrm{disrupt}}$ would be $m\gtrsim 2\times 10^{9}\,\mathrm{M_{\odot}}$, or 20 times larger than the median stellar mass of the UDGs (cf. Fig.~\ref{fig:stelmassdist}). We reach the same conclusion as \citet{vandokkum15}, namely that they must be highly centrally dark-matter dominated to survive down to such distances. The more compact dwarfs are observed down to a smaller radius of at least $D\simeq 100$ kpc, qualitatively consistent with the expectation for a system that is more tightly bound. Following the same argument, these also have to be centrally dark-matter dominated, with an estimated dark matter fraction of $\gtrsim$65\% within two times the effective radius of a galaxy with $M_{\star}\sim 10^{8}\,\mathrm{M_{\odot}}$ and $\mathrm{r_{eff}\simeq 1.0}$ kpc. Based on this comparison, we cannot rule out that UDGs have similarly massive dark-matter sub-haloes as the more compact dwarfs. To confirm and improve upon these estimates, one would require a study of their internal kinematics with extensive spectroscopic follow-up.

A key question is whether UDGs could have been more compact galaxies before they were accreted by a cluster (or by a lower-mass halo), and thus would have been part of the same parent population as the more compact dwarfs we study in Sect.~\ref{sec:comparisondwarfs}. Because galaxies fall into a cluster on different orbits, some galaxies of this population may have increased in size through tidal heating by interactions with the cluster or group centre, leading to UDGs with the size distribution presented in Fig.~\ref{fig:sizedist}. If UDGs belonged to the same parent population as the more compact dwarfs, before falling into the cluster, that would mean that these galaxies were able to hold on to a large amount of dark matter while undergoing the severe interactions with the cluster centres that are required to tidally heat the stellar component. Since tidal interactions with the cluster cores are also expected to have a strong effect on the surrounding dark matter subhaloes \citep[e.g.][]{limousin09}, it is questionable whether this scenario is feasible. 

An alternative scenario to consider is that the UDGs already had spatially extended stellar populations when they were accreted by the clusters. This would not only explain their high dark-matter fractions, but also their relatively high abundance compared to more compact dwarfs (with sizes $0.5<\mathrm{r_{eff}}<1.0$ kpc), and their radial distribution up to large clustercentric radii. 

In this case the differences in the properties of dwarf galaxies are most likely caused by variations in the relevant baryonic processes around the time of their formation. The colours of the UDGs and dwarfs suggest very old stellar populations, and hence that most of the star formation occurred at early times. The baryonic physics related to galaxy- and star-formation is complex in general, but is poorly understood in the early Universe where circumstances were very different from those in the local Universe. In particular, little is known about the formation of the first generation of stars from metal-poor (pristine) gas. This gas cools very inefficiently, and may fragment rather than all cool towards the centre of the halo. The resulting stellar population would have a more extended morphology, and retain this if gas gets expelled due to processes such as supernova explosions of the earliest stellar populations. How reionization proceeds, may also affect star formation in low-mass haloes in the early Universe \citep[e.g.][]{sawala16}, resulting in a strong dependence of the radial distribution of stars with the formation time with respect to the epoch of reionization. Alternatively, the recent study by \citet{amorisco16} suggests that the abundance of UDGs can be naturally explained by a simple model of disk formation, in which the sizes of galaxies are set by the spin of their haloes.

With our current analysis it is not possible to distinguish between these scenarios, but we foresee several avenues that can provide further insight. First, different models and simulations that aim to explain the origin of UDGs can be tested against the observed photometric properties such as the size distribution and radial profile within the clusters. But also a dedicated spectroscopic follow-up study could provide useful constraints, for example by studies of individual galaxies using spectroscopy to study their metal content and mean stellar ages \citep[e.g.][]{guerou15}. And finally, a large and dedicated field survey may help constrain their average volume density in the local Universe \citep{dalcanton97}, and the typical environments they live in outside of clusters.

\section{Summary and outlook}\label{sec:conclusion}
This paper provides a quantitative approach to identify UDGs in cluster data, and to study their properties. This is an important step towards understanding their origin and subsequent evolution in the cluster potential, since it allows for a more direct comparison with models and simulations. The main conclusions of this observational study can be summarized as follows.

\begin{itemize}
\item We introduce a reproducible method to robustly detect UDGs in clusters with photometric data, in a way that is quantitative as it does not rely on a by-eye inspection of galaxy images.
\item The selected UDGs have $g-r$ colours that are fully consistent with the cluster red sequence, extended towards fainter galaxies. Due to the degeneracy between age and metallicity, we cannot constrain either of these well. The median stellar mass of the sample, however, does depend less sensitively on these assumptions ($M_{\star}\simeq 10^{8}\,\mathrm{M_{\odot}}$).
\item Each of the eight $0.044<z<0.063$ galaxy clusters we study contains a significant population of UDGs. This indicates that the galaxies that were recently found in the Coma (and Virgo) clusters seem to be a general cluster phenomenon. Moreover, the relation between the number of UDGs and the host halo masses, which span an order of magnitude in this sample, is almost linear ($N\propto M^{0.93\pm0.16}$). 
\item The size distribution at a given surface brightness is so steep that the largest galaxies only constitute a minor part of the total UDG population in the clusters. The best-fitting size distribution of UDGs, at a given surface brightness, falls off as $\mathrm{n\, [dex^{-1}]\propto r_{eff}^{-3.4\pm 0.2}}$.
\item The projected radial distribution of UDGs is very steep in the cluster outskirts (steeper than NFW), and flat near the cluster centres (compared to NFW). An Einasto profile, with a relatively large curvature parameter of $\alpha_\mathrm{EIN}=0.92^{+0.08}_{-0.18}$, provides a good fit to the data. We note that this curvature parameter is much larger than for Einasto profiles that are generally used to describe the total mass distribution in dark matter haloes.
\item We also compare the radial number density distribution to that of more massive blue and red galaxies in these clusters (which were studied in vdB15). The UDGs seem to follow the total stellar-mass-weighted distribution of quiescent galaxies in the outskirts, indicating a dynamically old cluster population. However, compared to this distribution, the UDGs have a much lower central density, which is consistent with zero in the central region of $\mathrm{r_{core}}=0.15\times R_{200}$ (corresponding to $\sim 300$ kpc). That UDGs can exist down to a clustercentric distance as small as $\sim 300$ kpc indicates that they must be highly centrally dark-matter dominated.
\end{itemize}

When comparing the properties of cluster UDGs to those of more compact galaxies with similar luminosities, the following remarks can be made:
\begin{itemize}
\item Although small dwarf galaxies appear numerous in cluster images, they have an overdensity compared to the field that is significantly lower than for UDGs. After statistically subtracting the contribution from fore- and background galaxies, dwarf galaxies with sizes $0.5<\mathrm{r_{eff}<1\,kpc}$ are only a factor $\sim2$ more abundant than the cluster UDGs (with $\mathrm{r_{eff}>1.5\,kpc}$) in the same luminosity range.
\item These compact dwarfs follow the same radial distribution as UDGs in the outskirts, but can exist at smaller cluster-centric radii of at least $\sim 100$ kpc. This is expected for these more compact systems, given that they are more tightly bound and therefore more resistive to tidal disruption processes happening close to the cluster centres. This comparison may indicate that the more compact dwarfs are hosted by similarly massive dark-matter haloes as the UDGs.
\end{itemize}

Given the quantitative approach followed in this work, the results can be used as direct test to any model that seeks to explain the origin of these UDGs. For hydrodynamical simulations such a study poses a challenge, given the large dynamic range between a dwarf galaxy and its host halo, and the high spatial resolution required to distinguish UDGs from smaller dwarf galaxies. However, even without going into complex modelling or resorting to such cosmological simulations, we already discussed some basic properties of these galaxies in Sect.~\ref{sec:discussion}. Interestingly, the considered scenarios should result in observable differences of their properties. Follow-up spectroscopic studies of both the UDGs and the more compact dwarfs in these clusters are thus an interesting avenue to provide tighter constraints on their origin and their dynamical evolution in the clusters.

\begin{acknowledgements}
We thank Pierre-Alain Duc, Sophia Lianou, Michelle Collins, and Daniel Kelson for insightful discussions, Amandine Le Brun for comments on the manuscript, and Ricardo Herbonnet for providing a script that we used to run the galaxy simulations. We further thank the referee, Roberto Abraham, for a report with suggestions that helped us to clarify the paper. A large part of the data processing has been performed on a machine provided by Monique Arnaud at CEA Saclay.

The research leading to these results has received funding from the European Research Council under the European Union's Seventh Framework Programme (FP7/2007-2013) / ERC grant agreement n$^{\circ}$ 340519. Based on observations obtained with MegaPrime/MegaCam, a joint project of CFHT and CEA/IRFU, at the Canada-France-Hawaii Telescope (CFHT) which is operated by the National Research Council (NRC) of Canada, the Institut National des Science de l'Univers of the Centre National de la Recherche Scientifique (CNRS) of France, and the University of Hawaii. This work is based in part on data products produced at Terapix available at the Canadian Astronomy Data Centre as part of the Canada-France-Hawaii Telescope Legacy Survey, a collaborative project of NRC and CNRS. 
\end{acknowledgements}

\bibliographystyle{aa} 
\bibliography{MasterRefs} 

\begin{appendix}

\section{Completeness correction}\label{sec:radcompletenesscorr}
\begin{figure}
\resizebox{\hsize}{!}{\includegraphics{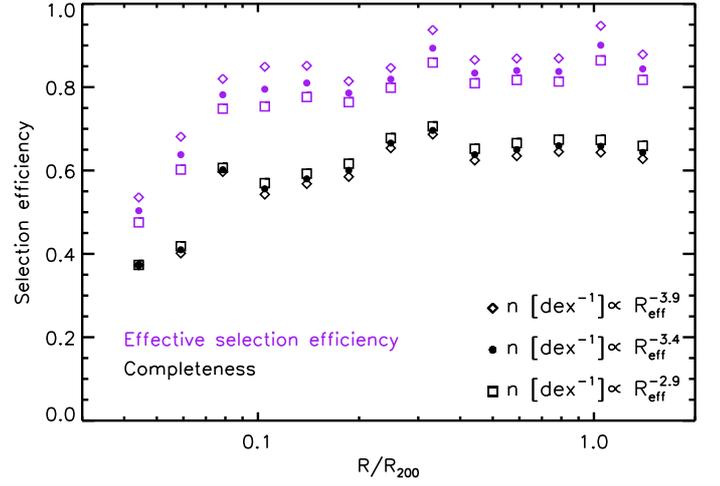}}
\caption{Radial completeness correction used. Two methods, which result primarily in a different normalisation of the radial profile. The central deficit is robust. }
\label{fig:radcompleteness}
\end{figure}
We measure the completeness of galaxies within our defined boundaries of parameter space, as a function of radial distance. These values depend slightly on the assumed intrinsic distribution, because a steep size distribution leads to many galaxies having effective sizes just above the 1.5 kpc boundary, and these may get lost due to the applied selection criteria. The different black symbols in Fig.~\ref{fig:radcompleteness} show the completeness, obtained from the simulations, for three different assumed size distributions. The effect of the size distribution on the completeness is small, considering the large statistical uncertainties on our measurements. Since this assumes that the size distribution is independent of radial distance, we test this assumption by measuring the size distribution of UDGs in the central R$<0.2\times$R$_{200}$, following the steps of Sect.~\ref{sec:sizedist}. The logarithmic slope we find, $-3.12\pm 0.31$, is similar to that of the full distribution within the (projected) virial radii, and therefore the filled black circles are our best estimate of the completeness. We find that the completeness varies by less than a factor of two from our central radial bin to the cluster outskirts. 

Rather than focussing on the mere completeness of our UDG detection/selection, it is important to also account for galaxies that are measured to be UDGs, but have intrinsic parameters just outside the selection box. We account for the size and surface-brightness distribution that we established in Sect.~\ref{sec:sizedist}. We then define an ``effective selection efficiency" as the number of objects that we identified as UDGs (whether correctly or not), divided over the total number of objects that should have been identified. We show this selection efficiency by the purple points in Fig.~\ref{fig:radcompleteness}. Note that the assumed slope has a different effect on this parameter. The steeper the slope, the more objects scatter into the UDG selection box, even though their sizes are just below 1.5 kpc, compared to objects that scatter downward in size (a type of \citet{eddington1913} bias). For a flat size distribution, the black and purple points would fall on top of each other, since there is no significant bias in the extracted parameters from \texttt{GALFIT}.
\end{appendix}

\end{document}